\documentclass[aps,prl,preprint,groupedaddress]{revtex4}
\usepackage{graphicx}
\begin{document}

\title{Spontaneously Induced Gravity:\\
From Rippled Dark Matter to Einstein Corpuscles}

\author{Aharon Davidson}
\email[Email: ]{davidson@bgu.ac.il}
\author{Ilya Gurwich}
\email[Email: ]{gurwichphys@gmail.com}
\affiliation{Physics department, Ben-Gurion University, Beer Sheva
84105, Israel}


\begin{abstract}
Suppose General Relativity, provocatively governed by a
dimensional coupling constant, is a spontaneously induced theory
of Gravity.
Invoking the Zee mechanism, we represent the reciprocal Newton
constant by a Brans Dicke scalar field, and let it damped oscillating
towards its General Relativistic VEV.
The corresponding cosmological evolution, in the Jordan frame,
averagely resembles the familiar dark radiation $\rightarrow$
dark matter $\rightarrow$ dark energy domination sequence.
The fingerprints of the theory are fine ripples, hopefully testable,
in the FRW scale factor; they die away at the strict General Relativity
limit.
Also derived is the spherically symmetric static configuration
associated with spontaneously induced General Relativity.
At the stiff scalar potential limit, the exterior Schwarzschild solution
is recovered.
However, due to level crossing at the would have been horizon,
it now connects with a novel dark core characterized by a locally
varying Newton constant.
The theory further predicts light Einstein-style gravitational
corpuscles (elementary particles?) which become point-like at the
GR-limit.\end{abstract}


\maketitle

Suppose General Relativity (GR) is spontaneously induced.
In which case, it makes sense to promote the (reciprocal) Newton constant
to the level of a scalar field $G^{-1}(x)$, and let it damped oscillate towards
its General Relativistic vacuum expectation value $<G ^{-1}(x)>=G^{-1}$.
The Zee realization of such an idea calls for a generalized (explicitly
broken conformal invariance) Brans-Dicke theory, where the a Brans-Dicke Lagrangian
\begin{equation}
	I=-\int \left(
	\phi R+\frac{\omega}{\phi}g^{\mu\nu}\phi_{,\mu}\phi_{,\nu}
	+V(\phi)+L_{matter}\right)\sqrt{-g}~d^{4}x 
\end{equation}
is supplemented by a tenable scalar potential $V(\phi)$.
For the sake of simplicity, one may momentarily switch off the matter
Lagrangian $L_{matter}$.
 
\medskip
Associated with the above action are the gravitational field equations
\begin{equation}
	\phi G_{\mu\nu}=-\phi_{;\mu\nu}
	+g_{\mu\nu}g^{\alpha\beta}\phi_{;\alpha\beta}
	+\frac{\omega}{\phi}\left(
	\frac{1}{2}g_{\mu\nu}
	g^{\alpha\beta}\phi_{,\alpha}\phi_{,\beta}
	-\phi_{,\mu}\phi_{,\nu}\right)
	+\frac{1}{2}g_{\mu\nu}V(\phi) ~,
\end{equation}
where $G_{\mu\nu}=R_{\mu\nu}-\frac{1}{2}g_{\mu\nu}R$
denotes the Einstein tensor.
Substituting the Ricci scalar curvature $R$ into the companion scalar 
field equation
\begin{equation}
	\omega\left(\frac{2}{\phi}\phi_{;\mu\nu}
	-\frac{1}{\phi^{2}}
	g^{\alpha\beta}\phi_{,\alpha}\phi_{,\beta}\right)
	= R +\frac{\partial V}{\partial \phi} ~,
\end{equation}
one finally arrives at
\begin{equation}
	g^{\mu\nu}\phi_{;\mu\nu}=\frac{1}{3+2\omega}
	\left(\phi\frac{\partial V}{\partial \phi}-2V\right)~.
\end{equation}
Counter intuitively, the emerging Klein-Gordon equation is governed
by the effective potential
\begin{equation}
	V_{eff}(\phi) =
	\frac{1}{3+2\omega}\int 
	\left(\phi\frac{\partial V}{\partial \phi}-2V\right) d\phi~.
\end{equation}
It is the absolute minimization of $V_{eff}(\phi)$, not of bare $V(\phi)$,
which sets the VEV of the theory.
Such a minimization procedure is generically accompanied by a
background cosmological constant
$\displaystyle{\Lambda\equiv -\frac{1}{4}\langle R
\rangle=\frac{1}{4}\frac{\partial V}{\partial \phi}}$, such that the constant
curvature vacuum solution is de-Sitter rather than Minkowski.
Up to an additive constant, a typical example is given by
\begin{equation}
	V(\phi)=6G\Omega^{2}(\phi-G^{-1})^{2}+2G\Lambda\phi^{2}
	\quad \Longleftrightarrow \quad
	V_{eff}(\phi)=2\Omega^{2}(\phi-G^{-1})^{2}
\end{equation}
It is crucial to notice that (i) The GR-limit is associated with
$\Omega\rightarrow\infty$, and (ii) The effective potential is
$\Lambda$-independent.

\medskip
To appreciate what is going on, here is a cosmological touch
for $\Lambda=0$ and $\omega=0$.
Within the framework of $k=0$ FRW cosmology, substituting one field
equation into the other leaves us with a second order master equation
for the Hubble constant	
\begin{equation}
	H\ddot{H}+3H^{2}\dot{H}-\frac{1}{2}\dot{H}^{2}
	+2\Omega^{2}H^{2}=0 ~.
\end{equation}

At early time ($\Omega t<<1$), we find
\begin{equation}
	H(t)\simeq\frac{1}{2t}+\frac{4\Omega p}{5}(\Omega t)^{2} ~,
	\quad
	G\phi(t)\simeq\frac{p}{\sqrt{\Omega t}}+
	\frac{4p^{2}}{5}\Omega t ~.
\end{equation}
Depending on the sign of $p$, weak initial gravity may be attractive
($p>0$) or repulsive ($p<0$).
It is only for the special case $p=0$ that the $\phi$-expansion starts
with $G\phi\simeq\frac{4}{5}\Omega^{2}t^{2}$, which implies strong,
necessarily attractive, initial gravity.
This special case is singled out on finiteness grounds when noticing
that $R\simeq 12\Omega^{2}$, rather than
$\displaystyle{R\simeq\frac{p}{\sqrt{t}}}$, as $t\rightarrow 0$.

\begin{figure}[ht]
	\includegraphics[scale=0.5]{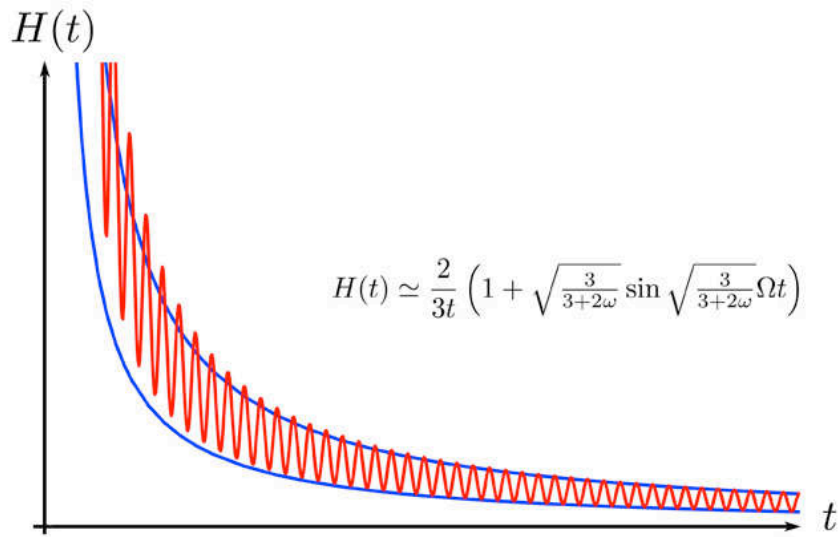}
	\caption{The Hubble constant $H(t)$ evolution for
	$\Lambda=0$. Time averaging over the ripples, justified
	for large $\Omega t$, resembles a General Relativistic
	matter dominated Universe.}
	\label{BDH} 
\end{figure}

\begin{figure}[ht]
	\includegraphics[scale=0.5]{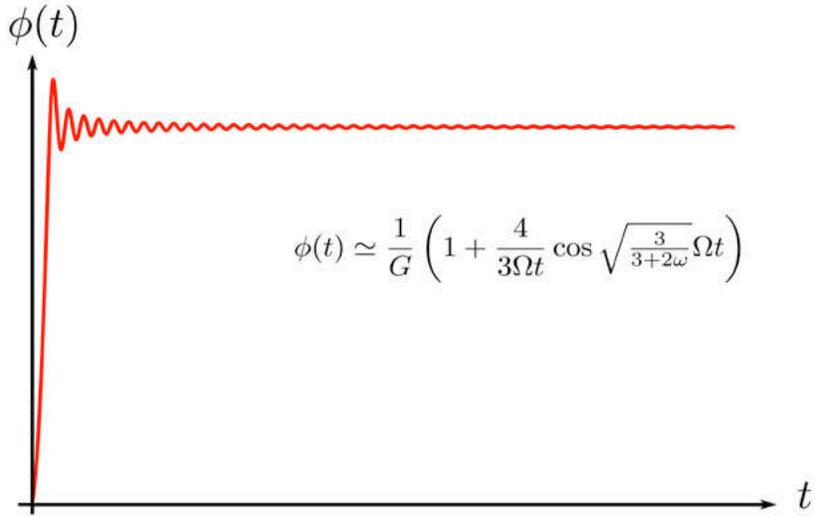}
	\caption{Approaching General relativity:
	The reciprocal Newton constant $G^{-1}(t)$ evolution for
	$\Lambda=0$ tells us, quite non-trivially, that the matter
	dominated Universe is in fact General Relativistic. }
	\label{BDphi} 
\end{figure}

\medskip
At later times ($\Omega t \gg 1$), asymptotically approaching GR, our
solution exhibits the particular dust dominated behavior modulated
by damped oscillations.
In particular, the corresponding FRW scale factor $a(t)$ can be integrated
to establish our main result
\begin{equation}
	a(t)\sim t^{2/3}\left(
	1+\frac{\sin{2\Omega t}}{3\Omega t}\right) ~,
\end{equation}
whose consequences are depicted in the Figs.\ref{BDH},\ref{BDphi}

\medskip
Two practical questions immediately arise:
(i) How large is this $\Omega$ frequency? and
(ii) Can one experimentally probe the rippled structure?
Obviously, following the Nyquist-Shannon theorem, if $\Omega^{-1}\ll\tau$,
$\tau$ denoting the Hubble time, one would not be able to tell the actual
FRW scale $a(t)$ from its coarse grained average.
Note in passing that the incorporation of a positive cosmological
constant $\Lambda$ would have a non-trivial effect on the cosmic ripples.
Whereas dark matter ripples are only $t^{-1}$-suppressed, dark energy
ripples are $\displaystyle{\exp(-\sqrt{\frac{3}{4\Lambda}}t)}$-suppressed.

\medskip
Another topic of interest is the static radially symmetric line element
\begin{equation}
	ds^{2}=-e^{\nu(r)}dt^{2}+e^{\lambda(r)}dr^{2}+
	r^{2}d\Omega^{2} ~.
\end{equation}
Based on the following gravitational and scalar field equations (keep
$\Lambda=0$, but allow arbitrary $\omega$)
\begin{eqnarray}
	&&\phi^{\prime\prime}-\frac{1}{2}(\nu^{\prime}+\lambda^{\prime})
	\left(\phi^{\prime}+\frac{2}{r}\phi\right)+
	\omega\frac{\phi^{\prime 2}}{\phi}=0 ~,\\
	&&\phi^{\prime\prime}+\frac{1}{2}(\nu^{\prime}-\lambda^{\prime})
	\left(\phi^{\prime}-\frac{2}{r}\phi\right)-\frac{2}{r^{2}}(1-e^{\lambda})
	\phi = \frac{2e^{\lambda}}{3+2\omega}
	\left\{\phi\frac{dV(\phi)}{d\phi}+
	\left(\omega-\frac{1}{2}\right)V(\phi)\right\} ~,\\
	&&\phi^{\prime\prime}+
	\left(\frac{\nu^{\prime}-\lambda^{\prime}}{2}+
	\frac{2}{r}\right)\phi^{\prime}=
	\frac{e^{\lambda}}{3+2\omega}
	\left\{\phi\frac{dV(\phi)}{d\phi}-2V(\phi)\right\} ~,
\end{eqnarray}
there are three basic issues need to be addressed:
(i) Deviations from Newton's force law,
(ii)  The fate of Schwarzschild black-hole away from GR, and
(iii) Novel non-GR configurations.

\medskip
Perturbing around the exact Schwarzschild solution, the leading order
equation can be solved analytically only for zero mass $M=0$; for
$M\neq 0$, a power series in $GM$ is in order.
Another constant of integration $\epsilon$, the would be scalar charge,
makes its appearance.
Unfortunately, due to length limitations, we are obliged to skip the
derivation, but once the dust settles down, Newton's constant is
supplemented by a Yukawa tail
\begin{equation}
	\phi(r) \simeq \frac{1}{G}
	\left(1+\frac{\epsilon}{r}e^{-\xi\Omega r}
	\left(1+GM\Omega
	\left(\frac{1}{\Omega r}-\xi 
	\left(e^{2\xi}\int^{\infty}_{2\xi}\frac{e^{-z}}{z}dz+\log \kappa r
	\right)\right)\right)\right) ~,
\end{equation}
where $\displaystyle{\xi\equiv\frac{2\sqrt{3}}{\sqrt{3+2\omega}}}$. 
By the same token, Newton's gravitational potential reads now
\begin{equation}
	\Phi(r)=-\frac{GM}{r}-\frac{\epsilon}{2r}e^{-\xi\Omega r}
	\left(1-GM\Omega
	\left(\frac{2}{\Omega r}+\xi
	\left(e^{2\xi}\int^{\infty}_{2\xi}\frac{e^{-z}}{z}dz+\log \kappa r
	\right)\right)\right) ~.
\end{equation}

\begin{figure}[ht]
	\includegraphics[scale=0.95]{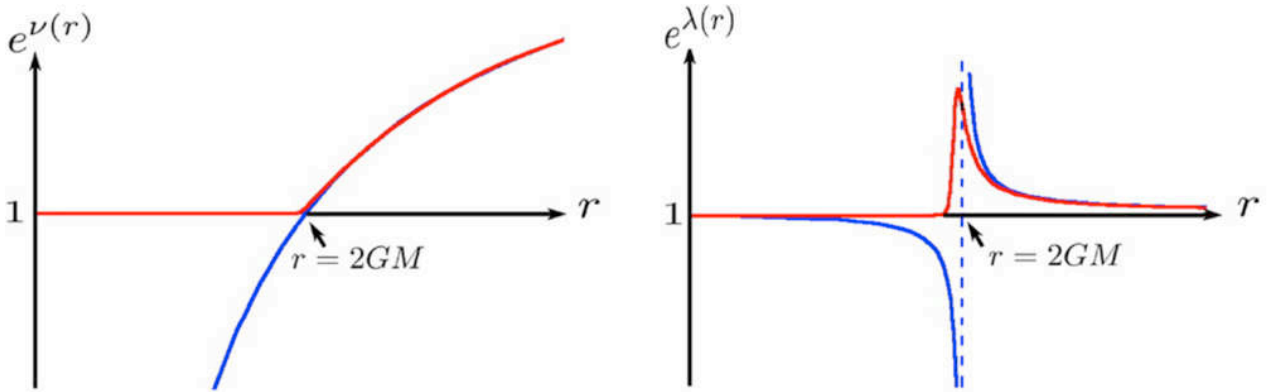}
	\caption{Horizon phase transition: At the stiff scalar
	potential limit $\Omega\rightarrow\infty$, the recovered exterior
	Schwarzschild solution (blue) connects with a novel interior core.}
	\label{M} 
\end{figure}

\begin{figure}[ht]
	\includegraphics[scale=1.1]{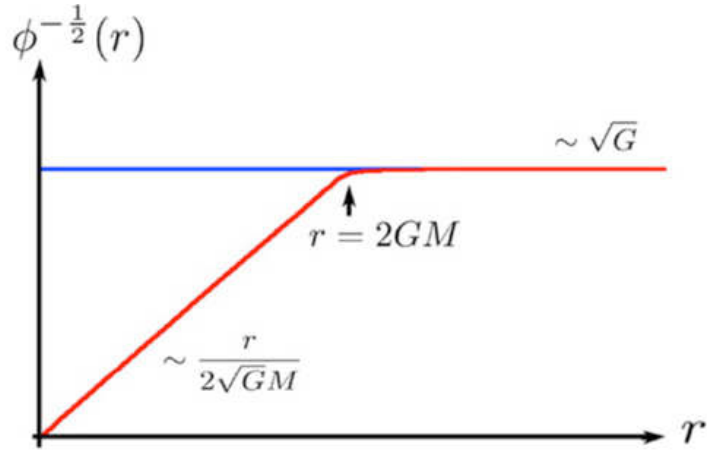}
	\caption{Radial dependence of the Newton 'constant'
	(the GR case is depicted in blue).}
	\label{phi} 
\end{figure}

\medskip
What happens when nearing the would have been Schwarzschild horizon?
At the stiff potential limit ($\Omega\rightarrow\infty$), the exterior
Schwarzschild solution (Newton constant $G$, mass $M$) is recovered.
However, due to a level crossing phenomenon which occurs at the
would have been horizon, it now connects with a novel dark core.
The latter is characterized by a varying Newton constant accompanied
by an exponentially small spatial volume element
\begin{equation}
	G_{in}(r)=\frac{r^2}{4GM^2} ~, \quad
	e^{\lambda(r)}_{in} \sim e^{\nu(r)}_{in} \sim  
	 \lim_{\delta\rightarrow 0}
	\left( \frac{r}{2GM} \right)^{\frac{1}{\delta^{2}}} ~.
\end{equation}
Altogether, as depicted by Figs.\ref{M},\ref{phi}, we face a horizon phase
transition, meaning a black star is formed.
Our theory further predicts light Einstein-style gravitational corpuscles
(elementary particles?) which become point-like at the GR-limit.
The universal size of such objects is ${\cal O}(\Omega^{-1})$.
\begin{figure}[ht]
	\includegraphics[scale=0.95]{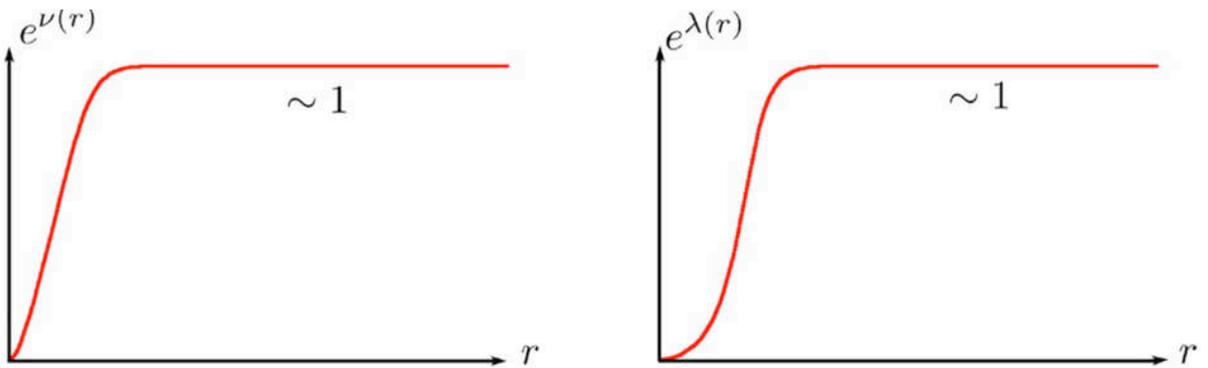}
	\caption{Einstein corpuscles: $M=0$ localized
	 configurations of universal radius ${\cal O}(\Omega^{-1})$.}
	\label{M=0} 
\end{figure}

\subsection{References:}
\begin{enumerate}
	\item A. Zee, Phys. Rev. Lett. 42, 417 (1979).
	\item A. Davidson, Class. Quant. Grav. 22, 1119 (2005).
	\item A. Einstein, Sit. Preus. Akad. (1919).
	\item A. Davidson and I. Gurwich (in preparation).
\end{enumerate}

\end{document}